\documentclass[12pt]{article}
\usepackage[english,german,french,polish]{babel}
\usepackage[T1]{fontenc}
\usepackage{amsfonts}

\begin{document}

\selectlanguage{english}

\baselineskip 0.75cm
\topmargin -0.6in
\oddsidemargin -0.1in

\let\ni=\noindent

\renewcommand{\thefootnote}{\fnsymbol{footnote}}

\pagestyle {plain}

\setcounter{page}{1}

\pagestyle{empty}

~~~

\begin{flushright}
IFT--05/19
\end{flushright}

{\large\centerline{\bf What if the LSND effect gets}}

{\large\centerline{\bf a considerably smaller amplitude?}}

\vspace{0.4cm}

{\centerline {\sc Wojciech Kr\'{o}likowski}}

\vspace{0.3cm}

{\centerline {\it Institute of Theoretical Physics, Warsaw University}}

{\centerline {\it Ho\.{z}a 69,~~PL--00--681 Warszawa, ~Poland}}

\vspace{0.6cm}

{\centerline{\bf Abstract}}

\vspace{0.2cm}

The possibility is considered that the verdict of the ongoing MiniBooNE neutrino experiment will 
favor neither of the contesting sides, stating in fact that the LSND effect with the original 
oscillation amplitude is not confirmed, but a new LSND effect with a considerably smaller amplitude
is found (or, at least, strongly suggested). Then, in the framework of neutrino oscillations, the presence in Nature of hypothetical light sterile 
neutrinos (mixing with three active neutrinos with a weak strength) will be unambiguously 
suggested by post-MiniBooNE neutrino data (unless the credible CPT invariance is seriously 
violated). In the new situation,  the 3+1 neutrino models may work all right. The same is true also for
simple 3+2 neutrino models.  For illustration of this potential developement, the simplest 3+2 
model is briefly discussed.

\vspace{0.2cm}

\ni PACS numbers: 12.15.Ff , 14.60.Pq , 12.15.Hh .

\vspace{0.6cm}

\ni August 2005 

\vfill\eject

~~~
\pagestyle {plain}

\setcounter{page}{1}

\vspace{0.2cm}

\ni {\bf 1. Introduction}

\vspace{0.2cm}

In these days, a lot of physicists are holding their breath and wait for the verdict of MiniBooNE neutrino experiment about the existence or nonexistence of the LSND effect. In the framework of  of neutrino oscillations, this possible effect is unambiguously connected with the presence in Nature of hypothetical light sterile neutrinos mixing with three active neutrinos with a weak strength (unless the credible CPT invariance is, in fact, seriously violated) [1]. 

Among different potential verdicts of the ongoing MiniBooNE neutrino experiment there is one which favors neither of the contesting sides, stating that the LSND effect with the original oscillation amplitude is not confirmed, but a new LSND effect with a considerably smaller amplitude is found (or, at least, strongly suggested).

If such a specific post-MiniBooNE situation is realized, the 3+1 neutrino models with one light sterile neutrino, estimated as disfavored by pre-MiniBooNE data (including the original LSND results) [2],  may work all right with a new, considerably weaker, strength of mixing between active and sterile neutrinos. Also, two simple 3+2 neutrino models with two light sterile neutrinos, described in Refs. [3] and [4], may be correct with a new, considerably weaker, mixing strength, though they are disfavored by pre-MiniBooNE data similarly as the 3+1 neutrino models. However,  the generic 3+2 models discussed with the use of statistical arguments [5] provide much better global fits to pre-MiniBooNE data than the 3+1 models. The 3+2 neutrino models may be relevant also in the specific 
post-MiniBooNE situation (where the oscillation amplitude in the new LSND effect is considerably smaller). 

Notice, however, the potential problem that light sterile neutrinos, displaying a weak but significant mixing with active neutrinos, are disfavored by thermodynamical equilibrium arguments if applied to neutrinos in the early Universe and confronted with astrophysical observations of helium and deuterium in the present Universe [6].The weak mixing strength discussed in Eq. (10) in the framework of our simplest 3+2 model [4] is "significant"\, in the sense of the above arguments. Certainly, a sufficiently weak mixing strength of active and sterile neutrinos would allow us to avoid the cosmological problem of sterile neutrinos. The question is as to whether such a mixing strength could be observed in the near future (the respective estimations are not obvious for us).

\vspace{0.2cm}

\ni {\bf 2. The simplest 3+2 model}

\vspace{0.2cm}

The simple 3+2 neutrino model from Ref. [4], being in a sense the simplest 3+2 model, is defined by the $5\times 5$ mixing matrix $U^{(5)} = \left( U^{(5)}_{\alpha i} \right) \;(\alpha = e, \mu, \tau, s, s'\;{\rm and}\; i=1, 2, 3,4,5)$ of the form 

\begin{eqnarray}
U^{(5)} & = & U^{(5)}(12) U^{(5)}(14,25) \nonumber \\
 & = & \left( \begin{array}{ccccc} 
c_{12}c_{14} & s_{12}c_{25} & 0 & c_{12}s_{14} & s_{12}s_{25} \\ 
-\frac{1}{\sqrt2}s_{12}c_{14}\;\;\, & \frac{1}{\sqrt2}c_{12}c_{25} & \frac{1}{\sqrt2} & -\frac{1}{\sqrt2}   s_{12} s_{14}\;\;\, & \frac{1}{\sqrt2}c_{12}s_{25} \\ 
\frac{1}{\sqrt2}s_{12}c_{14} & -\frac{1}{\sqrt2}c_{12}c_{25}\;\;\, & \frac{1}{\sqrt2} & \frac{1}{\sqrt2} s_{12}s_{14} & -\frac{1}{\sqrt2}c_{12}s_{25}\;\;\, \\
-s_{14}\;\;\, & 0 & 0 & c_{14} & 0 \\ 0 & -s_{25}\;\;\, & 0 & 0 & c_{25} 
\end{array} \right)  \,,
\end{eqnarray}

\ni where 

\vspace{-0.2cm}

\begin{eqnarray}
U^{(5)}(12)\;\;\;\; & = & \left( \begin{array}{ccccc} 
c_{12} & s_{12} & 0 & 0 & 0 \\ 
\!-\frac{1}{\sqrt2}s_{12} & \frac{1}{\sqrt2}c_{12} & \frac{1}{\sqrt2}& 0 & 0 \\ 
\frac{1}{\sqrt2}s_{12} & \!\!-\frac{1}{\sqrt2}c_{12} & \frac{1}{\sqrt2} & 0 & 0 \\
0 & 0 & 0 & 1 & 0 \\ 
0 & 0 & 0 & 0 & 1 \end{array}\right) \,, \nonumber \\
U^{(5)}(14,25) & = & \;\;\left( \begin{array}{ccccc} 
c_{14} & 0 & 0 & s_{14} & 0 \\ 
0 & c_{25} & 0 & 0 & s_{25} \\ 0 & 0 & 1 & 0 & 0 \\
\!\!-s_{14} & 0 & 0 & c_{14} & 0 \\ 0 & \!\!-s_{25} & 0 & 0 & c_{25} \end{array} \right) \,.
\end{eqnarray}

\ni In this model, the unitary transformation

\begin{equation}
\nu_\alpha  = \sum_i U^{(5)}_{\alpha i}\, \nu_i 
\end{equation}

\ni  holds between the flavor neutrinos $\nu_e \,,\, \nu_\mu \,,\, \nu_\tau\,,\, \nu_s\,,\, \nu_{s'} $ and mass neutrinos $\nu_1, \nu_2 , \nu_3, \nu_4, \nu_5 $. Here, the assumptions $c^2_{14} \gg s^2_{14} > 0$ and $c^2_{25} \gg s^2_{25} > 0$ imply the weak but still considerable mixing of two sterile neutrinos $\nu_s\,,\, \nu_{s'}$ with three active neutrinos $\nu_e \,,\, \nu_\mu \,,\, \nu_\tau$. Besides, there is no mixing between $\nu_s$ and $\nu_{s'}$.

With the notation

\begin{equation} 
x_{j i} \equiv 1.27 \frac{\Delta m^2_{j i} L}{E}\; , \; \Delta m^2_{j i} \equiv m^2_j - m^2_i
\end{equation}

\ni  ($\Delta m^2_{j i}$, $L$ and $E$ are measured in eV$^2$, km and GeV, respectively), we obtain after calculations [4] the following neutrino oscillation probabilities  in the Chooz reactor and LSND accelerator experiments:

\begin{equation}
P(\bar{\nu}_e \rightarrow \bar{\nu}_e)_{\rm Chooz} \simeq 1 - 2\left(c^2_{12}s^2_{14} + s^2_{12} 
s^2_{25} \right) \sim 1\; ({\rm experimentally})
\end{equation}

\ni and

\begin{equation}
P(\bar{\nu}_\mu \rightarrow \bar{\nu}_e)_{\rm LSND}  \simeq  2c^2_{12} s^2_{12}\left[ s^2_{14} s^2_{25} \sin^2 (x_{54})_{\rm LSND} + \frac{1}{2}\left( s^2_{14} - s^2_{25}\right)^2 \right] \,,
\end{equation}

\ni when

\begin{eqnarray}
m^2_1 < m^2_2 \ll m^2_3 \ll m^2_4 < m^2_5\;\; & , & \Delta m^2_{21} \ll \Delta m^2_{54} \ll \Delta m^2_{41} \,, \nonumber \\
\left( x_{31}\right)_{\rm Chooz} \simeq \left( x_{31}\right)_{\rm atm} = O(\pi/2) & , & \;\;\;\left( x_{54}\right)_{\rm LSND} = O(\pi/2) \,.
\end{eqnarray}

\ni Recall that in our model $s_{13} = 0$ (and, consequently, CP is not violated in neutrino oscillations). If $s^2_{14} \simeq s^2_{25}$, Eqs. (5) and (6) are simplified:

\begin{equation} 
P(\bar{\nu}_e \rightarrow \bar{\nu}_e)_{\rm Chooz} \simeq  1 - 2 s^2_{14} = \cos 2\theta_{14} \sim 1\; ({\rm experimentally}) 
\end{equation}

\ni and

\begin{equation}
P(\bar{\nu}_\mu \rightarrow \bar{\nu}_e)_{\rm LSND} \simeq 2c^2_{12} s^2_{12} s^4_{14} \sin^2 (x_{54})_{\rm LSND} \,.
\end{equation}

\ni Thus, in this model, the third squared-mass scale, needed to produce the new LSND effect, is given by $\Delta m^2_{54} \equiv m^2_5 - m^2_4$. 

For example, if the new LSND effect exists with the amplitude of the order $10^{-3}$, then from Eq. (9)

\begin{equation} 
s^2_{14} \sim \left(\frac{10^{-3}}{2c^2_{12}s^2_{12}}\right)^{1/2} \sim 0.0482 \,,
\end{equation}

\ni where $2c^2_{12}s^2_{12} \sim 0.431$, as follows from the solar neutrino experiments ($s^2_{12} \sim 0.314$ [6]). In this case, $\theta_{14} \simeq \theta_{25} \sim 12.7^\circ  $ giving $P(\bar{\nu}_e \rightarrow \bar{\nu}_e)_{\rm Chooz} \simeq \cos 2\theta_{14} \sim 0.904 $, what equals 1 with the deviation of 9.6\% and so is nearly at the (Chooz) experimental edge.

\vfill\eject
\vspace{0.2cm}

\ni {\bf 3. The simplest 3+1 model}

\vspace{0.2cm}

The simplest 3+1 neutrino model ({\it cf.} Appendix in Ref. [4]) can be obtained from our 3+2 model by putting $s_{25} \rightarrow 0$ and $(x_{41})_{\rm LSND} = O(\pi/2)$. Then, after calculations [4],

\begin{equation} 
P(\bar{\nu}_e \rightarrow \bar{\nu}_e)_{\rm Chooz}  \simeq 1 - 2 c^2_{12}s^2_{14} \sim 1 \;({\rm experimentally})
\end{equation} 

\ni and

\begin{equation} 
P(\bar{\nu}_\mu \rightarrow \bar{\nu}_e)_{\rm LSND} \simeq 2 c^2_{12} s^2_{12} s^4_{14} \sin^2 (x_{41})_{\rm LSND}.
\end{equation}

\ni In Eq. (12), the amplitude is the same as before, but the third squared-mass scale, producing the new LSND effect, is provided by $\Delta m^2_{41} \equiv m^2_4 - m^2_1$ in place of the previous $\Delta m^2_{54} \equiv m^2_5 - m^2_4$.

For example, the expression for $s^2_{14}$ following from Eq. (12) gets the same form (10) as before, if the new LSND effect exists with the amplitude of the same order $10^{-3}$. Then, using Eq. (11) we obtain $P(\bar{\nu}_e \rightarrow \bar{\nu}_e)_{\rm Chooz} \sim 0.934$, what is equal to 1 with the deviation of 6.6\%, smaller than the previous deviation.

\vspace{0.2cm}

\ni {\bf 4. Conclusion}

\vspace{0.2cm}

Thus, both our neutrino models, the simplest 3+2 and the simplest 3+1, may explain a new LSND effect  with an amplitude considerably smaller than that in the original LSND effect if, of course, the potential new effect is really found in MiniBooNE experiment or its possible continuation. Notice that, even at present, such a new LSND effect is not excluded by the pre-MiniBooNE data other than the original LSND results, when its amplitude is of the order $O(10^{-3})$ or smaller.

\vfill\eject

~~~~

\vspace{0.8cm}

{\centerline{\bf References}}

\vspace{0.5cm}

{\everypar={\hangindent=0.7truecm}
\parindent=0pt\frenchspacing

{\everypar={\hangindent=0.7truecm}
\parindent=0pt\frenchspacing

[1]~{\it Cf.} M.H~Shaevitz, {\tt hep--ex/0407027}.

\vspace{0.2cm}

[2]~{\it Cf. e.g.} M. Maltoni, T. Schwetz, M.A.~Tortola and J.W. Valle, {\it Nucl. Phys.} {\bf B 643}, 321 (2002).

\vspace{0.2cm}

[3]~W. Kr\'{o}likowski, {\it Acta Phys. Pol.} {\bf B 35}, 1675 (2004) [{\tt hep--ph/0402183}].

\vspace{0.2cm}

[4]~W. Kr\'{o}likowski, {\tt hep--ph/0506099}.

\vspace{0.2cm}

[5]~M. Sorel, J. Conrad and M. Shaevitz, {\tt hep--ph/0305255}.

\vspace{0.2cm}

[6]~{\it Cf. e.g.} S.H. Hansen {\it et al}, {\tt astro--ph/0105385}, where other arguments connected with anisotrophy of cosmic microwave background radiation were also used to confirm the bound $N_\nu \sim 3$ for the effective number of all neutrino species, following from the previous arguments.

\vspace{0.2cm}

[7]~G.L. Fogli, E. Lisi, A. Marrone and A. Palazzo, {\tt hep--ph/0506083}.

\vfill\eject

\end{document}